\documentclass[superscriptaddress, aps, prx,  longbibliography]{revtex4-2} %The [superscriptaddress] makes the \address as an affiliation working
\usepackage{graphicx} % Required for inserting images
\usepackage{array}
\usepackage{siunitx}
\usepackage[section]{placeins}
\usepackage{hyperref}
\usepackage{natbib}%dit MOET! natbib zijn,  met bibtex/biblatex gaat alles fout en kan je voor de rest van je leven niet meer compileren
\usepackage[version=4]{mhchem}
\begin{document}

\preprint{APS/123-QED}
% \title{\textbf{Observation of Thermal Motion of a \\ Sub-kHz Mechanical Resonator Passively Cooled to 6 mK} 
%\title{\textbf{Thermal Motion of a Sub-kHz \\Mechanical Resonator to below 10 mK.} 
% }% 
\title{\textbf{A Sub-kHz Mechanical Resonator Passively Cooled to 6 mK}} 

\author{Loek van Everdingen}\email{These authors share first authorship and contributed equally.}
\affiliation{%
 Leiden Institute of Physics, Leiden University, P.O. Box 9504, 2300 RA Leiden, The Netherlands.\\
}%

\author{Jaimy Plugge}\email{These authors share first authorship and contributed equally.}
\affiliation{%
 Leiden Institute of Physics, Leiden University, P.O. Box 9504, 2300 RA Leiden, The Netherlands.\\
}%

\author{Tim M. Fuchs} 
\affiliation{%
 Leiden Institute of Physics, Leiden University, P.O. Box 9504, 2300 RA Leiden, The Netherlands.\\
}%
\affiliation{%
School of Physics and Astronomy, University of Southampton, SO17 1BJ, Southampton, United Kingdom.\\
}%

\author{Guido L. van de Stolpe}
\affiliation{%
 Leiden Institute of Physics, Leiden University, P.O. Box 9504, 2300 RA Leiden, The Netherlands.\\
}%
\affiliation{E. L. Ginzton Laboratory, Stanford University, 348 Via Pueblo, Stanford, California, United States of America}

\author{Dalal Benali}
\affiliation{%
 Leiden Institute of Physics, Leiden University, P.O. Box 9504, 2300 RA Leiden, The Netherlands.\\
}%

\author{Thijmen de Jong}
\affiliation{%
 Leiden Institute of Physics, Leiden University, P.O. Box 9504, 2300 RA Leiden, The Netherlands.\\
}%

\author{Jasper Bijl}
\affiliation{%
 Leiden Institute of Physics, Leiden University, P.O. Box 9504, 2300 RA Leiden, The Netherlands.\\
}%

\author{Wim A. Bosch}
\affiliation{HDL, Hightech Development Leiden, Leiden, The Netherlands, HDL@freedom.nl}

%\affiliation{%
% Leiden Institute of Physics, Leiden University, P.O. Box 9504, 2300 RA Leiden, The Netherlands.\\
%}%

\author{Tjerk H. Oosterkamp}
 \email{Contact author: oosterkamp@physics.leidenuniv.nl}
\affiliation{%
 Leiden Institute of Physics, Leiden University, P.O. Box 9504, 2300 RA Leiden, The Netherlands.\\
}%

%  \homepage{http://www.Second.institution.edu/~Charlie.Author}
% \affiliation{
%  First affiliation for this author
% }%
% \affiliation{
%  second institution for this author
% \affiliation{%
%  Authors' institution and/or address\\
%  This line break forced with \textbackslash\textbackslash
% }%


\date{\today}% It is always \today, today,

\begin{abstract}

Highly coherent mechanical resonators are invaluable to ultrasensitive detection techniques by enabling detection of small forces. Studying mechanical resonators in a thermal equilibrium state at millikelvin temperatures provides a promising path to increase their coherence time. Here, we passively cool a $\SI{700}{Hz}$ massive ($\SI{1.5}{ng}$) mechanical cantilever down to $6.1 (4) \si{mK}$ by means of nuclear demagnetization, as confirmed by detecting its thermal motion via a lock-in based detection scheme. At the lowest temperatures the thermal motion of the resonator is still clearly distinguishable from the background noise. Our data analysis confirms that at these temperatures the motion is still thermally distributed. These results pave the way for passive cooling  low-frequency resonators to the sub-milllikelvin regime, which would enable new tests of quantum mechanics and advances in ultrasensitive force detection.

% \begin{description}
% Secondary publications and information retrieval purposes.
% You may use the \texttt{description} environment to structure your abstract;
% use the optional argument of the \verb+\item+ command to give the category of each item. 
% \end{description}
\end{abstract}

\maketitle

%\tableofcontents

Nanomechanical resonators play a vital role in fundamental physics, as they enable detection of small forces~\cite{fogliano_ultrasensitive_2021, heritier_nanoladder_2018, monteiro_accel_2020}. These forces play a role in topics such as nanoscale magnetic resonance imaging~\cite{rugar_single_2004, degen_nanoscale_2009, grob_mrfm_2019, fischer_spin_2019}, solid state physics\cite{bossoni_vortex_2014, wang_phase_2010, sellies_single_2023} and small-scale measurements of gravity~\cite{westphal_measurement_2021, fuchs_measuring_2024}. Additionally, large mass mechanical resonators play a role in studying the quantum-to-classical boundary by placing bounds on various effective wavefunction collapse models, such as Diosí-Penrose models and Continuous Spontaneous Localization (CSL)~\cite{Carlesso_CSL_2022,vinante_upper_2016,bassi_collapse_2013}.  Aforementioned experiments require mechanical resonators with favorable properties, such as high Q-factors~\cite{eichler_highQ_2022, seis_groundstate_2022}, low force noise~\cite{heritier_nanoladder_2018, janse_currentbounds_2024,  gisler_softclamped_2022} and good displacement sensitivity~\cite{lee_nanoscale_2023, debonis_displacement_2018}.
These properties tend to improve at lower resonator temperature and consequently experiments are generally carried out at cryogenic temperatures~\cite{stowe_attonewton_1997}. 

To reach such low temperatures, both active and passive forms of cooling are explored. Active cooling requires continuous driving of a mode, such as feedback cooling~\cite{zoepfl_kerr_2023} or sideband cooling~\cite{youssefi_squeezed_2023}, and has been employed to cool mechanical resonators in a wide range of frequencies into their ground state~\cite{teufel_sideband_2011, delic_cooling_2020, guo_feedback_2019, seis_groundstate_2022}. However, it is undesirable in experiments probing quantum systems, as it disturbs a system from mechanical equilibrium via the external drive. On the contrary, passive cooling leaves a mechanical system in thermal equilibrium, once cold. It was first used to cool GHz systems to their ground state at the operation temperature of a dilution refrigerator~\cite{Oconnel_groundstate_2010}. A widely-applied form of passive cooling is nuclear demagnetization, which has been used to cool microelectronics below $\SI{1}{mK}$~\cite{sarsby_nanoelectronics_2020, batey_microkelvin_2013}, and was recently employed to cool a 15 MHz device into its motional ground state~\cite{cattiaux_macroscopic_2021}. 

However, bringing sub-kHz mechanical probes to sub-mK temperatures while leaving them in thermal equilibrium remains a formidable challenge~\cite{pickett_european_2018}. If achieved, it has the potential to improve the Q-factor of highly-coherent mechanical resonators by mitigating mechanical loss channels~\cite{maillet_damping_2023, van_heck_magnetic_2023}. Mechanical sensors with a  high Q-factor and low-force noise find application in the aforementioned experiments in ultrasensitive detection, which would consequently benefit from a decrease in temperature.

In this work, we apply nuclear demagnetization to cool a $\SI{1.5}{ng}$ mechanical resonator to $\SI{6.1 (4)}{\milli \kelvin}$. By application of a lock-in detection scheme that tracks the resonator energy, we are able to verify the thermal equilibrium nature of its state through the direct observation of Boltzmann energy statistics. To our knowledge, this is the first observation of the equilibrium motion of a massive mechanical resonator in the Hz- to kHz-regime below the $\SI{20}{\milli \kelvin}$ base temperature of conventional dilution refrigerators.\\

% \section{\label{sec:meth}Methodology}
\subsection{\label{sec:meth}Experimental setup}

\begin{figure*}
    \centering
    \includegraphics[width=\textwidth]{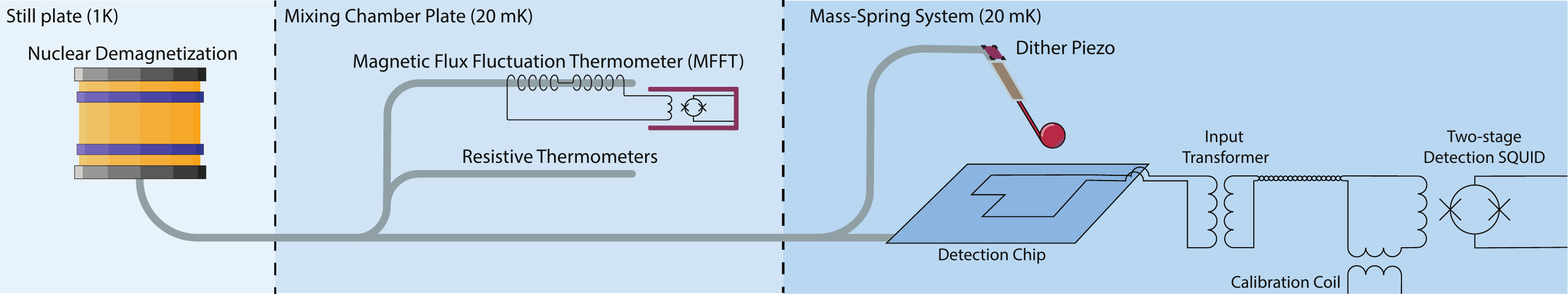}
    \caption{ Schematic illustration of the setup used for this work. A thermally isolated silver wire is linked to the nuclear demagnetization stage and connects different parts of the experiment, notably the cantilever, detection chip and input transformer to a temperature below the base temperature of the dilution refrigerator at $\SI{20}{\milli\kelvin}$. The lower right corner of the schematic shows the circuit that is used for the detection and calibration of the cantilever motion. The position inside the dilution refrigerator of different parts is indicated. In the center part the MFFT is illustrated inside its lead shielding (purple). The rightmost part shows the cantilever suspended above the detection circuit.}
    \label{fig:setup_schematic}
\end{figure*}

The experimental setup used in this work was originally designed for Magnetic Resonance Force Microscopy (MRFM) at low temperatures (see Fig. ~\ref{fig:setup_schematic}). A soft silicon cantilever~\cite{chui_mass-loaded_1995} with a magnetic tip (\ce{Nd2Fe14B} sphere, 7.3 $\mu\textrm{m}$ diameter) is suspended above a pickup loop in a pulse tube cryogen-free dilution refrigerator (Leiden Cryogenics CF-1200). To minimize mechanical vibrations, the experiment is mounted on a mass-spring suspension system hanging inside the dilution refrigerator. The cantilever exhibits a resonance frequency $\omega_0 \approx 2\pi \times \SI{700}{Hz}$ and a spring constant of $\SI{26}{\mu Nm^{-1}}$. Magnetic flux induced in the pickup loop generates a current in a SQUID input coil of a two-stage readout SQUID (Magnicon NC-1 Integrated Two-Stage Current Sensor), and hence a voltage signal. Precise positioning of the cantilever is achieved using a piezomotor system comprising three independently controllable spindles actuated by slip-stick piezomotors. Oscillations of the cantilever can be driven electrically through a dither piezo that excites the cantilever base. The amplitude of the cantilever oscillations can be calibrated through a calibration coil which excites the cantilever magnetically, and which is located between the pickup loop and the readout SQUID input coil. The cantilever frequency varies at different positions due to local forces acting on the cantilever. The local forces cause a change in the effective stiffness $k_{\textrm{eff}}$, which causes variations in the effective resonance frequency $\omega_{\textrm{eff}}$. \\

To achieve cooling of the cantilever below the base temperature of the dilution refrigerator ($\sim \SI{20}{\milli \kelvin}$), we thermally anchor both the cantilever and the detection chip to a \ce{PrNi5} nuclear demagnetization stage using a silver wire (contrary to previous work where only the cantilever was connected ~\cite{van_heck_magnetic_2023}). Crucially, the silver strip is routed along the mass-spring system using thermally isolating clamps, so that it is mechanically, but not thermally, connected to each mass. This allows the wire to reach sub-millikelvin temperatures (the mass-spring system is thermalized at $\sim \SI{20}{\milli \kelvin}$), while retaining optimal vibration isolation ~\cite{van_heck_magnetic_2023}. Thermal isolation between the detection chip and the sample holder is achieved by placing the detection chip on top of a machined Macor\textsuperscript{\textregistered} plate.

In order to accurately determine the temperature of the environment, three thermometers are connected to the silver wire, which forms a thermal link to the cantilever. We use two resistance thermometers for temperatures between $\SI{15}{\milli \kelvin}$ and $\SI{250}{\milli\kelvin}$ (from Hightech Development Leiden, HDL) , and a Magnetic Flux Fluctuation Thermometer (MFFT) for lower temperatures, which induces minimal heat dissipation ~\cite{fleischmann_noise_2020}. The MFFT consists of a gradiometric coil wound around the silver wire that picks up magnetic fluctuations induced by Johnson-Nyquist noise, which is measured using a second Magnicon SQUID \cite{van_heck_magnetic_2023}. Details on the thermometer calibration procedure can be found in the supplementary material~\cite{supplementary}.\\%~\ref{App:SQUID_thermometry}.\\

We directly infer the cantilever temperature by detecting its thermal motion through the AC magnetic flux signal induced in the SQUID pickup loop \cite{usenko_mrfm_2011}. This measurement of the cantilever motion is minimally invasive, with added dissipation in principle only limited by the coupling between the cantilever and the detector. We measure the flux-to-voltage conversion parameter $\kappa$ by driving the cantilever with an oscillating test flux generated by a calibration coil (see Fig. \ref{fig:setup_schematic}) and measuring the resulting flux through the pickup coil \cite{van_heck_magnetic_2023}. Note that here we only consider magnetic interactions, neglecting  possible electrostatic contributions (e.g. due to residual charges on the cantilever tip). A detailed discussion on the calibration procedure can be found in the supplementary material~\cite{supplementary}. \\

This work presents experiments in two different cooldown cycles of the dilution refrigerator, for which the respective measurement parameters are presented in table~\ref{tab:meas_runs}. In each cycle, we start by varying the coupling between the detection circuit and the cantilever before turning on the nuclear demagnetization. The resulting conversion parameters $\kappa$ are given in appendix~\ref{app:data}. In both cooldown cycles, the cantilever is more than $\SI{10}{\mu m}$  away from the surface of the detection chip while measuring. \\

% \section{\label{sec:res}Results}

\subsection{Determination of the Cantilever Temperature through its Thermal Motion}

\begin{figure*}
    \centering
    \includegraphics[width= \textwidth]{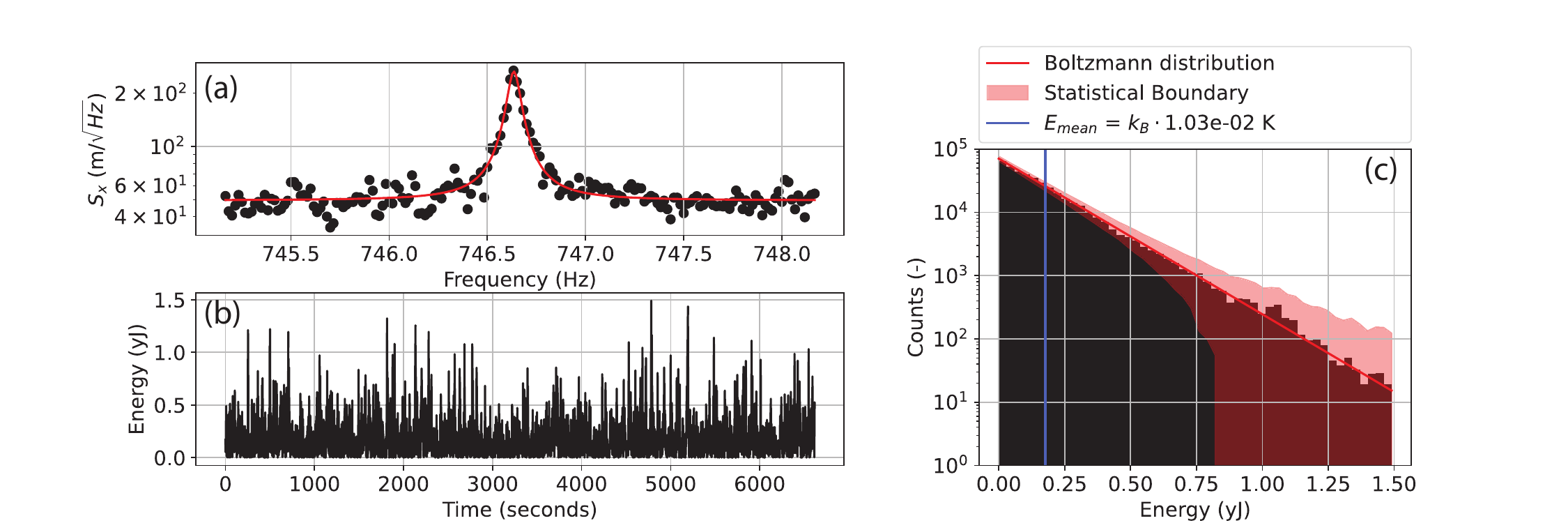}
    \caption{The cantilever temperature is determined from the readout SQUID signal after post-processing. (a) The resonance frequency of the cantilever is determined through a frequency sweep driven by the piezo. This is done through a Lorentzian fit over the range where the peak stands out above the detector noise. (b) The (undriven) thermal motion of the cantilever. A digital lock-in amplifier is used at the determined resonance frequency on the SQUID time signal to obtain the cantilever amplitude as a function of time. (c) After plotting the energy in a histogram, the cantilever temperature is obtained from the mean energy and crosschecked through the slope of the energy distribution. The red shaded area indicates one standard deviation in the number of counts per bin.}
    \label{fig:data_analysis_temperature}
\end{figure*}

To extract the cantilever temperature, we apply a lock-in-based detection scheme that enables real-time tracking of the cantilever energy (here executed in post-processing, see supplementary material~\cite{supplementary} for details on the analysis). This type of analysis was first implemented by Golokolenov \textit{et al.}~\cite{golokolenov_mesoscopic_2023} Such an energy time trace facilitates immediate insight into the resonator thermodynamics (see Fig. \ref{fig:data_analysis_temperature}b) which we use here to verify the thermally limited nature of its motion (Fig. \ref{fig:data_analysis_temperature}c). That is to say, it allows to check for occasional large mechanical disturbances that are not filtered by the mass-spring system. This method offers a clear and intuitive way to analyze the cantilever signal, and can be considered complementary to frequency domain methods \cite{van_heck_magnetic_2023}. 

The thermal motion of the cantilever is given by $\left< x \right>^2 = \textrm{k}_B T / k $, where $\left< x \right>^2$ is the time-averaged amplitude of the cantilever displacement, and $k$ is the cantilever stiffness (by invoking the equipartition theorem ~\cite{wagenaar_resonance_2013}). The cantilever frequency can vary due to local forces, such as Meissner repulsion induced by the superconducting structures of the readout circuit or due to magnetic interactions with surface spins \cite{van_heck_magnetic_2023, denhaan_dissipation_2015}.  We measure the effective resonance frequency $\omega_{\textrm{eff}}$ by exciting the cantilever magnetically through the calibration coil or mechanically through the piezo and fitting a Lorentzian to the Fourier transform of the SQUID signal (10 minutes of integrated signal is shown in Fig.~\ref{fig:data_analysis_temperature}a). \\

The (digital) lock-in amplifier is applied to the measured time signal from the readout SQUID at resonance frequency $\omega_{\mathrm{eff}}$. We set the lock-in bandwidth to $\SI{1}{Hz}$, approximately two times the bandwidth over which the thermal motion of the resonator stands out above the detection noise (see Fig. \ref{fig:data_analysis_temperature}a).  The correlation time of the cantilever is $\tau = 2\frac{Q}{\omega_0}$, which determines the timescale at which its energy decays. A value for $\tau$ can be determined by sweeping cantilever excitations around the resonance frequency. Typically we find $\tau \approx \SI{7}{\second}$. Hence, two hours of data, as presented in figure~\ref{fig:data_analysis_temperature}b, contains about 1028 independent measurements of the cantilever energy, from which we infer the cantilever temperature by taking the mean. To verify the thermal nature of the data, we compare a histogram to the Boltzmann distribution, which scales as: $\sim \exp{\left(\frac{E}{-k_{\textrm{B}}T}\right)}$ (Fig. \ref{fig:data_analysis_temperature}c). A cutoff is visible (i.e. the number of counts drops abruptly from 25 to zero) at high energy, because the sample rate of the lock-in is much faster than the rate at which the cantilever energy varies, determined by $\tau$. External (mechanical) excitations of the cantilever motion would cause the histogram in Fig. \ref{fig:data_analysis_temperature}c to deviate from a Boltzmann distribution. \\

To crosscheck the temperature, we define an expression for the standard deviation $\delta n$ of the distribution. Given the number of independent measurements in a bin is equal to $t_{\textrm{bin}}/\tau$, with $t_{\textrm{bin}}$ the bin duration, we expect the relative variation to be:

\begin{equation}
    \frac{\delta n}{n_{\textrm{bin}}} = \frac{1}{\sqrt{t_{\textrm{bin}}/\tau}}\, , 
\end{equation}

 for a bin that contains $n_{\textrm{bin}}$ counts. $95 \%$ of the data are expected to fall within an interval of $\pm 2\textrm{ }\delta n$ around Boltzmann distribution, which is confirmed by our analysis (see Fig. \ref{fig:data_analysis_temperature}c). By fitting the slope of the distribution we extract a cantilever temperature of $ 10.3 (2) \si{\milli \kelvin}$ (where the value between brackets denotes the uncertainty on the last digit).\\

\subsection{Thermal Motion below 10 mK}

The results in this work concern a comparison between two independent datasets of the cantilever temperature. These datasets are used to discuss the factors that cause saturation of the cantilever temperature and factors that cause offsets in the calibration procedure for the cantilever amplitude.\\

Each dataset consists of a measurement of thermal fluctuations of the cantilever when changing the current through the nuclear demagnetization coil. The magnetic field of the nuclear demagnetization coil is first ramped up to $\SI{2}{\tesla}$. The energy released during this process is thermalized to the dilution refrigerator through an aluminum heat conductance switch. After the field has reached $\SI{2}{\tesla}$ and the demagnetization stage has thermalized, the heat conductance switch is flipped to minimize thermal conductivity. Then the field is reduced stepwise, providing cooling power to the silver wire through the demagnetization of the \ce{PrNi5}. At the lowest magnetic field, the cantilever and detection chip are allowed to thermalize for at least 12 hours, while the amplitude of the cantilever resonance is being monitored and the MFFT acquires temperature data. When no further decrease in the measured cantilever PSD is observed, the current through the nuclear demagnetization coil is increased stepwise, while monitoring the cantilever amplitude. After each step, we wait for the MFFT temperature to reach a stable value after which two hours of data are acquired. \\

%This experiment is repeated in two different cooldown cycles of the dilution refrigerator, for which the respective measurement parameters are presented in table~\ref{tab:meas_runs}. In each cycle, we vary the coupling between the detection circuit and the cantilever before turning on the nuclear demagnetization. The resulting conversion parameters $\kappa$ are given in appendix~\ref{App:Energy_coupling}. In both datasets, the cantilever is more than $\SI{10}{\mu m}$  away from the surface of the detection chip while measuring. \\

\begin{figure}
    \centering
    \includegraphics[width=\textwidth]{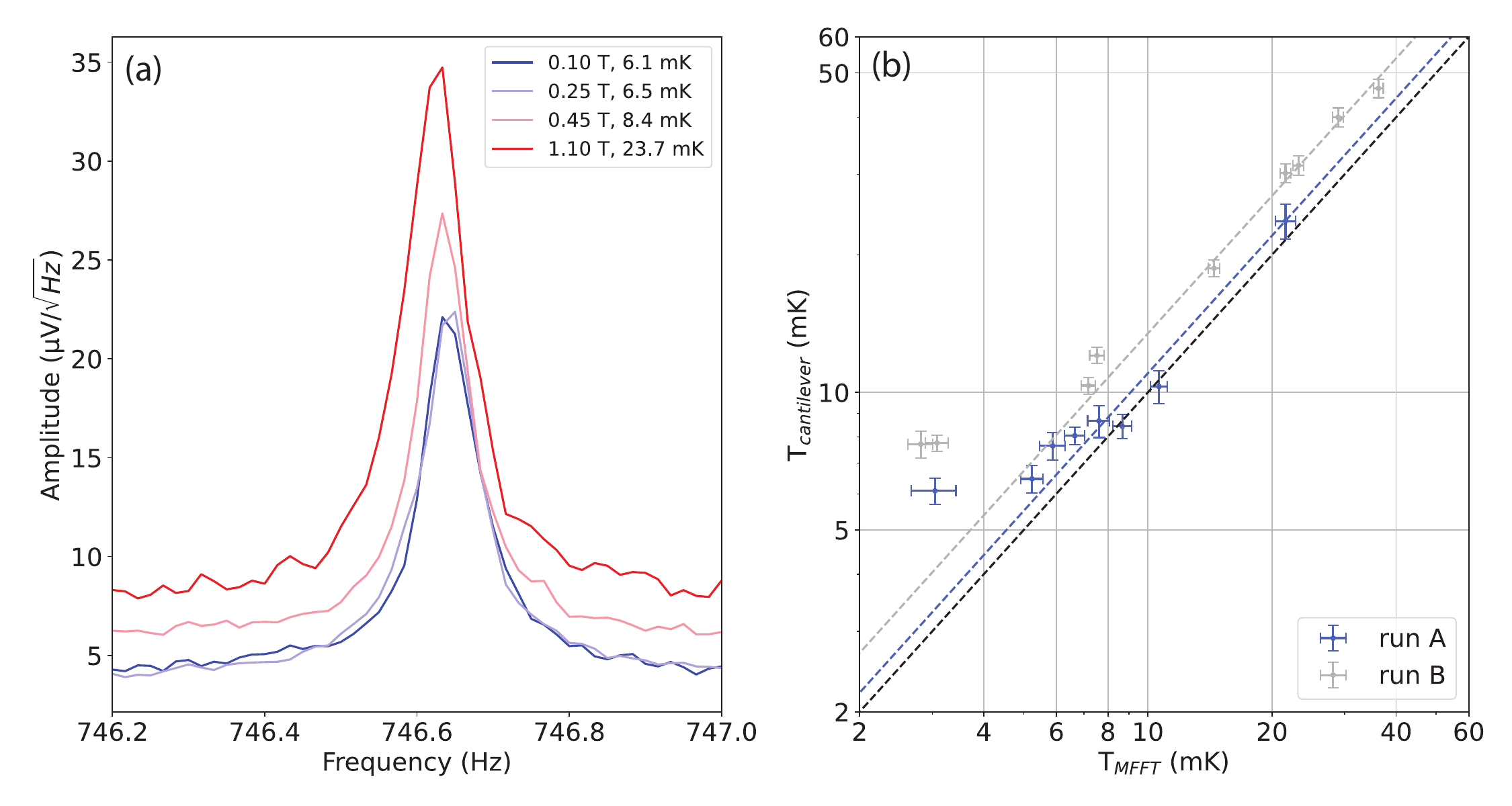}
    \caption{Temperature determined from the cantilever thermal motion versus bath temperature. (a) shows the thermal motion as observed in the power spectral density during measurement run A at various magnetic fields in the nuclear demagnetization stage. In (b) the cantilever temperature $T_{\textrm{cantilever}}$ is plotted against the MFFT temperature $T_{\textrm{MFFT}}$ during run A and B. The dashed lines indicate $T_{\textrm{cantilever}} = c T_{\textrm{MFFT}}$ corresponding to run A and B. The black dashed line indicates  $c = 1$.}
    \label{fig:different_runs}
\end{figure}

The thermal motion measured in dataset A is plotted as a PSD in figure~\ref{fig:different_runs}a to be able to observe a decrease in amplitude at lower temperatures. The final temperature is extracted from the corresponding histograms, which can be found in appendix~\ref{app:histograms}. During all measurements, the thermal fluctuations of the cantilever can clearly be distinguished from the background detection noise. The measured thermal fluctuations of the cantilever, plotted as a cantilever temperature  $T_{\textrm{cantilever}}$, are plotted against the MFFT temperature $T_{\textrm{MFFT}}$ in figure~\ref{fig:different_runs}b.  Horizontal error bars indicate the standard deviation of the temperature measurements from different spectra throughout this time interval. If the cantilever is in thermal equilibrium, its temperature is expected to fluctuate around an average value $\bar{T}_{\textrm{avg}}$. The standard deviation of these fluctuations over a time interval $t_{\textrm{meas}}$ is described using $\Delta T = \sqrt{\tau/t_{\textrm{meas}}}\bar{T}_{\textrm{avg}}$ around the average temperature.\\

\begin{table}
\centering
\caption{Parameters of the two different measurement runs. \textbf{T$_{\textrm{final}}$} is the lowest temperature measured through the cantilever thermal motion during a run, \textbf{T$_{\textrm{MFFT}}$} the lowest temperature measured through the MFFT.}
\label{tab:meas_runs}
\resizebox{0.48\textwidth}{!}{%
\begin{tabular}{|l|l|l|l|l|l|l|}
\hline
\textbf{Run} &   \textbf{Frequency} (Hz) & \textbf{$Q$} (-) & \textbf{$\kappa$} (m/V) & \textbf{T$_{\textrm{final}}$} (mK) & \textbf{T$_{\textrm{MFFT}}$} (mK)\\ \hline
A            &  746.6             & 13200 (100) &  $9.6 (6)\cdot 10^{-6}$              & $6.1 (4)$                    & $3.4 (4)$                         \\ \hline
B            &  669.7              & 15400 (400)  &   $ 1.9 (2) \cdot 10^{-5}$ &       $7.7 (4) $ & $3.1 (2)$                         \\ \hline
\end{tabular}%
}
\end{table}

Throughout both datasets, we observe that ${T_{\textrm{cantilever}} > T_{\textrm{MFFT}}}$ to within the uncertainty of $T_{\textrm{cantilever}}$. At the lowest temperatures, $T_{\textrm{cantilever}}$ is saturated and does not decrease further, at a saturation temperature of $\SI{6.1(4)}{\milli\kelvin}$ and $\SI{7.7(4)}{\milli\kelvin}$ for run A and B, respectively. Even at the lowest temperatures, the cantilever motion is still following a Boltzmann distribution, hinting at a thermally distributed origin of the saturation.\\

To quantify the deviation between $T_{\textrm{cantilever}}$ and $T_{\textrm{MFFT}}$, we fit $T_{\textrm{cantilever}} = cT_{\textrm{MFFT}}$ for the data with $T_{\textrm{MFFT}}>\SI{8}{\milli \kelvin}$. The uncertainty in the fit parameter is estimated from the square root of its variance. However, the uncertainty is dominated by the limited amount of fitting points, which is not reflected in this value. For the data in run A, the value for $c = 1.07(3)$, close to $1$, which is expected when the cantilever and MFFT are well thermalized to each other. However in run B, $c$ is equal to $1.33(3)$. As no changes were made to the silver wire or cantilever between datasets A and B, we suspect the deviation in $c$ to be the result of inaccuracies in the displacement calibration of the cantilever. A possible origin is the presence of electrostatic forces driving the cantilever. The calibration only takes into account forces due to generated magnetic fields. However, if there are any residual charges on the cantilever tip, electrostatic forces can act as an additional  drive of unknown sign on the cantilever due to parasitic capacitances. Appendix~\ref{App:offset} discusses potential inaccuracies in the displacement calibration in more depth. If the calibration in run B was lower such that $c = 1$, it would result in a saturation temperature around $\SI{6}{mK}$, similar to dataset A. \\

We deem it unlikely that the saturation of the cantilever temperature around $\SI{6}{mK}$ is due to external vibrations. Vibrations would result in day-night variations of the cantilever temperature. We do not observe such variations, as is visible in appendix~\ref{app:data}. We fit the functional ${T_{\textrm{cantilever}} = \left(T_{\textrm{MFFT}}^n + T_0^n\right)^{1/n}}$ to the data of run A. The  free parameters  are the saturation temperature $T_0$ and coefficient $n$. The value of $n$ can be used to determine the limiting thermal resistance~\cite{usenko_squid_2010, pobell_matter_2007}. This yields $n = 4 (2)$ and $T_0 = 6  (1) \si{\milli \kelvin}$. This is consistent with a thermal resistance coupled in through the detection chip. A hypothesis for this saturations is heating due to the thermal dissipation of the DC bias voltage in the SQUID detection circuit.\\

\subsection{\label{sec:conclusion}Discussion}

\DeclareSIUnit{\sqrthertz}{\sqrt{\unit{\hertz}}}

Through the use of nuclear demagnetization in combination with a vibration isolation system, we demonstrated passive cooling of a nanomechanical cantilever to temperatures below $\SI{10}{mK}$, measuring its equilibrium motion at temperatures down to $6.1 (4) \si{\milli \kelvin}$. This result paves the way for improvements in the detection sensitivity of nanomechanical sensors, by improving their force-noise and Q-factor through equilibrium cooling. Additionally, it constitues a steps towards future tests of wavefunction collapse models such as CSL~\cite{Carlesso_CSL_2022}.\\

The success of these experiments is dependent on the force noise of the cantilever for which we can make an estimate using the method of Stowe \textit{et al.} ~\cite{stowe_attonewton_1997}. Our current cantilever has a resonance frequency of approximately $2\pi \cdot\SI{700}{Hz}$, effective mass is $m_{\textrm{eff}} \approx \SI{1.5}{ng}$ and $Q \approx 14000$. At $\SI{6.1}{mK}$ this yields $\sqrt{S_F} \approx  \SI[per-mode = symbol]{3.9e-19}{\newton\per\sqrthertz}$. This is on the same order as the best efforts to optimize force noise in the low-kHz regime ~\cite{ranjit_zeptonewton_2016, mamin_subatto_2001} and two orders of magnitude higher than the state of the art for clamped mechanical resonators~\cite{debonis_displacement_2018, fogliano_ultrasensitive_2021}.\\

This figure can be improved by a number of changes to the experiment. The experiments performed by van Heck \textit{et al.}, were carried out with the exact same cantilever, which then had $Q \approx 40000$. It is unclear what caused the deterioration of its Q-factor. A higher Q-factor can  be achieved by switching to nanoladder cantilevers~\cite{heritier_nanoladder_2018}. A nuclear demagnetisation stage using \ce{PrNi5} can theoretically reach a minimal temperature of $\SI{0.5}{\milli\kelvin}$. By combining  $Q \approx 40000$ with a cantilever temperature of $\SI{0.5}{\milli\kelvin}$ we can potentially achieve $\SI[per-mode = symbol]{6.8e-20}{\newton\per\sqrthertz}$. \\

There are two issues to overcome in order to cool a cantilever to $\SI{0.5}{mK}$. Firstly, one has to remove any mechanisms that currently cause saturation of the cantilever temperature. Potential origins are heating caused by the SQUID detection circuit and free electron spins on the detection chip that are excited by spurious magnetic field fluctuations. 
Secondly, one can improve the techniques to lower the demagnetization temperature by either increasing the cooling power of the experiment or by reducing the heat input through conduction or mechanical vibrations. This is possible through a second nuclear demagnetization system, utilizing nuclear spins in copper,  that is precooled using the existing nuclear demagnetization coil. \\  

Finally, for future experiments, it is important to reduce uncertainty in the calibration procedure. This requires a way to mitigate the effect of parasitic capacitances throughout the magnetic calibration or by reducing any remaining charges on the cantilever, making it less susceptible to the electrostatic driving force. \\

\begin{acknowledgments}
We would like to thank Maria Luisa Mattana for her valuable feedback on the manuscript. 

TO and LvE acknowledge funding from the Netherlands Science Organisation (NWO
grant OCENW.GROOT.2019.088) and from two Quantum Delta National Growth Fund grants.

\end{acknowledgments}

\appendix

\newpage

\section{Combined MFFT and Cantilever Temperature Data}\label{app:data}

\begin{figure}
    \centering
    \includegraphics[width= 0.99\textwidth]{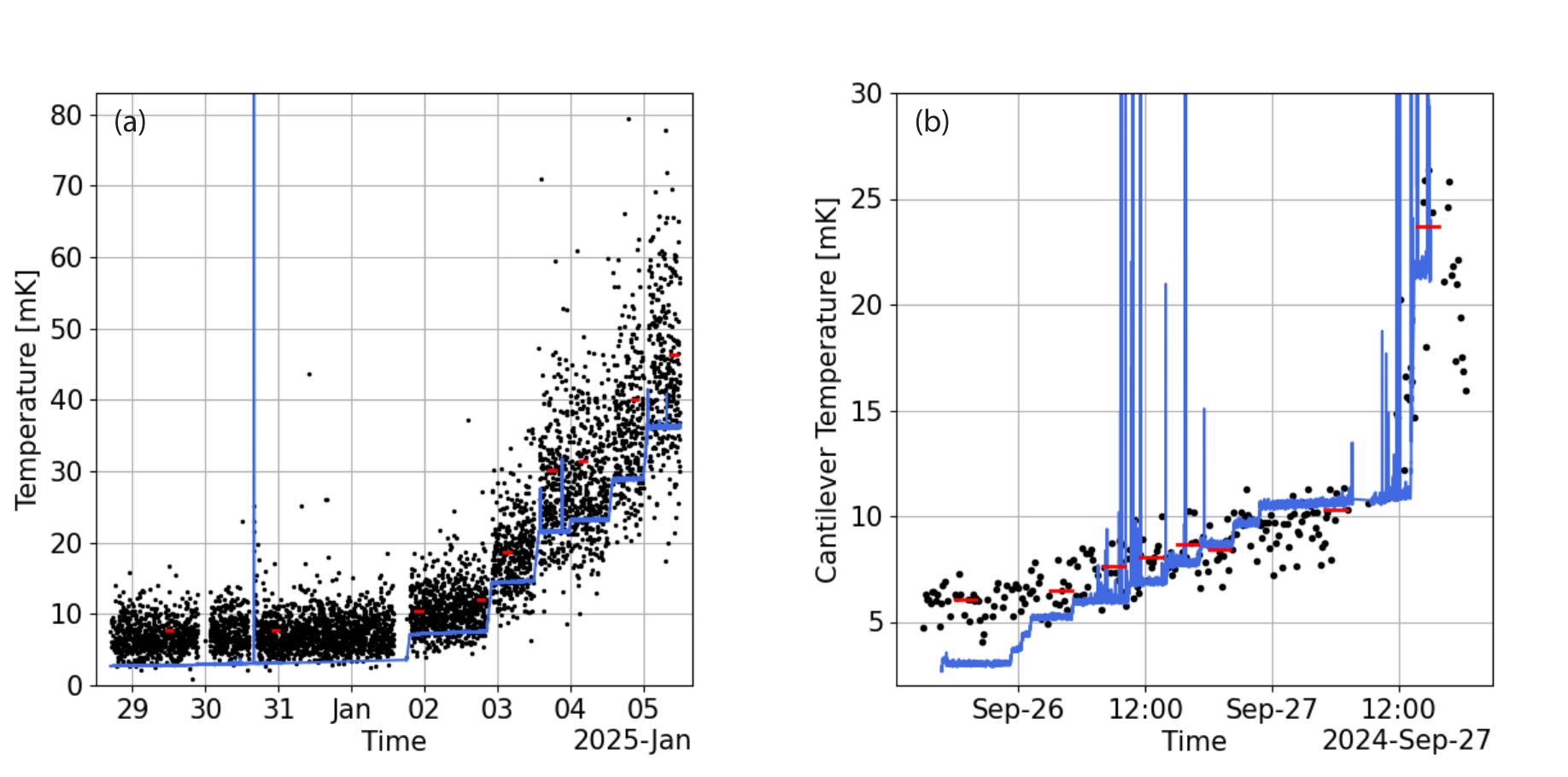} %

    \caption{Temperature steps measured during run A (a) and B (b). The temperature of the MFFT is plotted in blue and the cantilever temperature in black. Red bars indicate the time segments used to calculate the datapoints in figure~\ref{fig:different_runs}  (b) of the main text.}%
    \label{fig:DataOverview}%
\end{figure}

The combined temperature measurement through the MFFT and the temperature measured from the cantilever motion are displayed in figure~\ref{fig:DataOverview}. In table~\ref{tab:overview} contains the parameters resulting from the displacement calibration (details outlined in the supplementary material~\cite{supplementary}). \\

In the MFFT data presented in figure~\ref{fig:different_runs}, a datapoint is added for the temperature obtained from the spectrum from all datafiles (respectively $\SI{16}{\second}$ and $\SI{61}{\second}$ of measurement during run A and run B) obtained during a run. The spikes in the temperature occur when a change is made to a DC offset voltage in the electronics of the MFFT SQUID. This does not change the IV-curve of the SQUID and resulting sensitivity for changes in magnetic flux. However, due to the jump in the output voltage, a spike appears in the temperature data after processing.\\

\begin{table*}
\centering
\caption{Parameters for the cantilever displacement calibration as measured for the different runs in this work. Sample rate refers to the sample rate of the detection SQUID, The paramaters $\beta$ and $\kappa$ are respectively the energy coupling and conversion parameter from the displacement calibration. The temperatures \textbf{T$_{\textrm{final}}$} and \textbf{T$_{\textrm{MFFT}}$} are the lowest temperatures measured on from the cantilever motion and through the MFFT respectively.}
\label{tab:overview}
\resizebox{1\textwidth}{!}{%
\begin{tabular}{|l|l|l|l|l|l|l|l|l|}
\hline
\textbf{Run} &  \textbf{Sample Rate} (Sa/s) & \textbf{Frequency} (Hz) & \textbf{$Q$} (-)  & \textbf{$\beta$} (-) & \textbf{$Q\beta^2$} (-) & \textbf{$\kappa$} (m/V) & \textbf{T$_{\textrm{final}}$} (mK) & \textbf{T$_{\textrm{MFFT}}$} (mK) \\ \hline
A            & 200000                & 746.6              & 13200 (120) & $1.8 (6)\cdot 10^{-3}$ & $ 4.4 (6)\cdot 10^{-2}$                   & $9.6 (6) \cdot 10^{-6} $    & $6.1 (4)$                    & $3.4 (4)$                 \\ \hline
B            & 50000            & 669.7              & 15400 (400)  & $1.0 (5)\cdot 10^{-3}$  & $1.6 (4)\cdot 10^{-2}$                       & $1.9 (2)\cdot 10^{-5}$        &  $7.7 (4)$ & $3.1 (2)$                 \\ \hline
\end{tabular}%
}
\end{table*}

\section{Thermal Distributions of Cantilever Motion}\label{app:histograms}

\begin{figure}
    \centering
    \includegraphics[width=\textwidth]{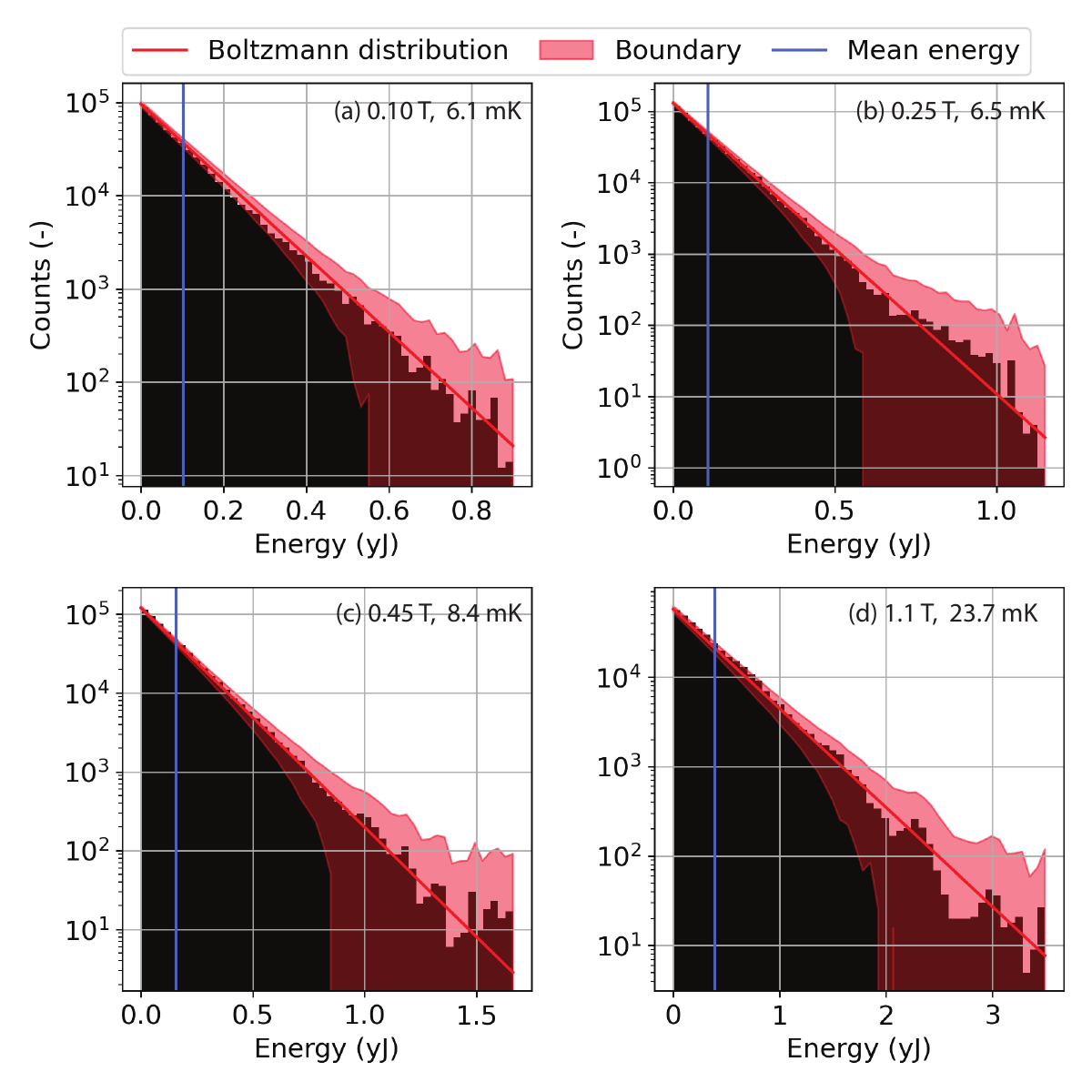}
    \caption{The histograms of the cantilever motion after applying a digital lock-in amplifier of the curves in figure~\ref{fig:different_runs} (a) in the main text. The red line marks the expected number of counts in each bin if they are drawn from a Boltzmann distribution with a width determined by the temperature as calculated from the mean energy. The red shaded area marks one standard deviation around this distribution. }
    \label{fig:histograms}
\end{figure}

In figure~\ref{fig:histograms} the histograms of the cantilever energy corresponding to the curves in figure~\ref{fig:different_runs}a are plotted. As indicated in the results section, variations in the cantilever energy occur over an interval determined by the inverse cantilever time constant $1/\tau$. This is much slower than the sample rate of the digital lock-in amplifier, which is 100 Sa/s. As a result, bins with less than $ \tau/10 * \SI{100}{Sa/s} \approx 70$ counts have little statistical significance.\\

\section{Offset in the Cantilever Temperature Conversion}\label{App:offset}

To obtain a quantified measurement of the cantilever temperature, several data processing steps are relevant. Here we discuss several issues that we identified in this procedure. Any inaccuracy can result in both a too low or too high figure for the cantilever temperature. We first present a potential systematic error due to deviations in the mass of the cantilever tip. Then we discuss the effect of electrostatic driving of the cantilever during the displacement calibration.\\

From the measured amplitude of the cantilever oscillations, the cantilever energy is calculated using a proportionality constant  $E \propto m_{\textrm{eff}}\omega^2$. The energy thus scales linearly with the effective cantilever mass, which is dominated by the mass of the spherical magnetic tip. We now discuss the uncertainty in the mass of this magnetic tip. The tip diameter was observed  to be $\SI{7.3}{\mu m}$, using a scanning electron microscope. Due to variations between different measurements, we estimate that this value is precise to about $8 \%$. Using the density of \ce{Nd2Fe14B} ($\rho = \SI{7450}{kg/m^3}$) we find that this translates to a mass of the tip of $\SI{1.51(35)}{ng}$. The cantilever mass is around $\SI{0.12}{ng}$. Taking into account the uncertainty in the mass of the tip, the uncertainty in the effective mass is found to be on the order of $23 \%$. This directly translates to a systematic uncertainty in the cantilever temperature, as $T_{\textrm{cantilever}} \propto m$.\\

\begin{figure}
    \centering
    \includegraphics[width=\textwidth]{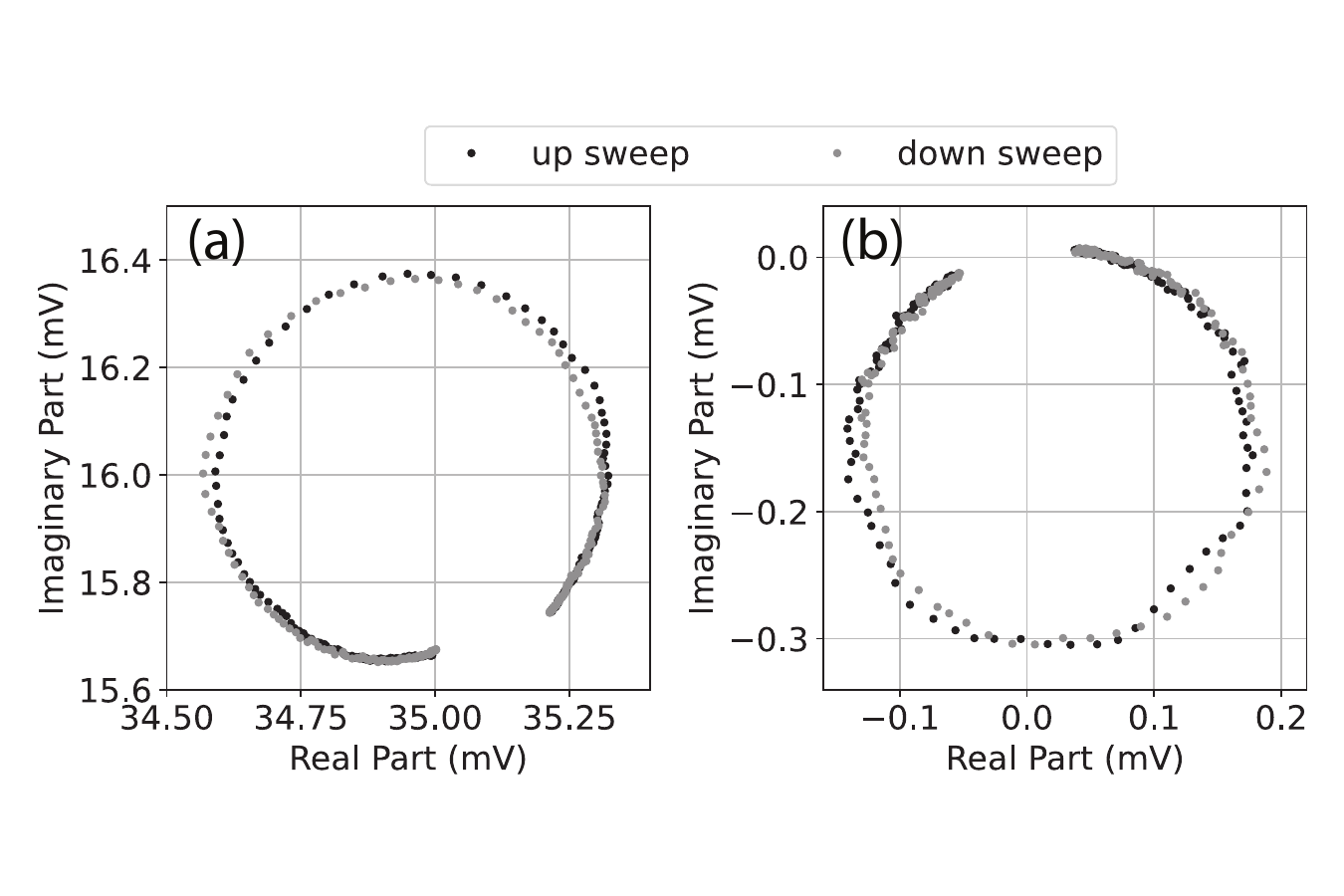}
    \caption{Two frequency sweeps through the calibration coil. In figure (a) the sweep is conducted in the normal configuration, with the input of the calibration coil connected to the output of the current source. In (b) is conducted to only allow for electrostatic driving of the cantilever. In both figures a circle is  visible that the cantilever is excited resonantly. However, in (b) there is no net offset of the circle from the origin, as there is no magnetic flux going through the calibration coil. Additionally, the circle is rotated with respect to the sweep conducted in (a).}
    \label{fig:electrostatic}
\end{figure}

We observed that it was possible to electrostatically drive the cantilever through the calibration coil. Figure~\ref{fig:electrostatic} shows a sweep through the calibration coil, during which the output of the lockin amplifier was grounded on two sides. This means that no current can run through the calibration coil and any effect is due to voltage changes. However, the cantilever is still resonantly driven during this sweep.  This indicates that the cantilever tip contains a non-zero electric charge, which is possibly acquired due to a touch of the detection chip while positioning. If electrostatic driving happens in parallel to inductive driving, the measured energy coupling $\beta^2$ can deviate from purely inductive driving.  If the resulting conversion parameter $\alpha$ is too high or too low, depends on the phase difference with which both driving mechanisms act on the cantilever. It was not possible to calculate the coupling off this effect using the procedure outlined in the supplementary material~\cite{supplementary}, as no offset flux is present during the SQUID measurement. This suggests there the SQUID signal is not affected through inductive driving during this measurement.\\

\nocite{chawner_lego_2019}
\nocite{woodcraft_low_2009}
\bibliography{apssamp}% Produces the bibliography via BibTeX.

\end{document}